\documentstyle[11pt,newpasp,twoside]{article}
\def\edcomment#1{\iffalse\marginpar{\raggedright\sl#1\/}\else\relax\fi}
\reversemarginpar

\begin{document}
\title{Kinematics and photometry as complementary tools in the
    study of barred galaxies}
\author{Juan Carlos Vega Beltr\'an, Peter Erwin, John Beckman}
\affil{IAC, C. Via L\'actea s/n, E-38200, La Laguna, Tenerife, Spain}
\author{Alessandro Pizzella, Enrico Maria Corsini, Francesco Bertola}
\affil{Universit\`a di Padova, Dipartimento di Astronomia,
Padova, Italy}
\author{Werner W. Zeilinger}
\affil{Institute of Astronomy, University of Vienna, 
Wien, Austria}


\section{UGC~10205: an edge-on barred galaxy}

It has been suggested that the peanut-shaped bulges seen in some edge-on
galaxies are due to the presence of a central bar (Bureau \& Freeman 1999;
Bureau \& Athanassoula 1999).  Although
bars cannot be detected photometrically in edge-on galaxies, Kuijken \&
Merrifield (1995) showed that a barred potential produces
strong kinematic signatures in the form of double-peaked line-of-sight
velocity distributions with a characteristic ``figure-of-eight'' variation
with radius.  As an example, in Fig.~1 (left panel) we can see two components
(fast-rotating and slow-rotating) that give the emission spectrum and the
velocity curve of the edge-on galaxy UGC~10205 a figure-of-eight shape (see
Vega et al. 1997).

\section{NGC 6221: Detecting Bar-Driven Shocks in the Gas}

The case of NGC 6221 is particularly interesting because the images show
numerous dust lanes within the bar, but whether these actually represent
shocks in the gas flow --- as predicted by numerous hydrodynamic simulations
--- is unclear. However, if we take into account the radial profiles of
[NII]/H$\alpha$ derived from long-slit spectra (see Vega et al. 1998), 
we can infer the presence of a pseudo-ring of shocked gas in
the inner $10^{\prime\prime}$ region.  If we compare the HST image with the
positions of these peaks (see Fig.~2, right panel), we see that the regions
of maximum [N II]/H$\alpha$ occur where the slits cross the two inner,
curving dust lanes which join the ``leading-edge'' lanes near the center of
this galaxy.  This suggests that the strongest shocks may occur in the two
curving lanes, rather than in the leading-edge lanes.
\begin{figure}
\includegraphics{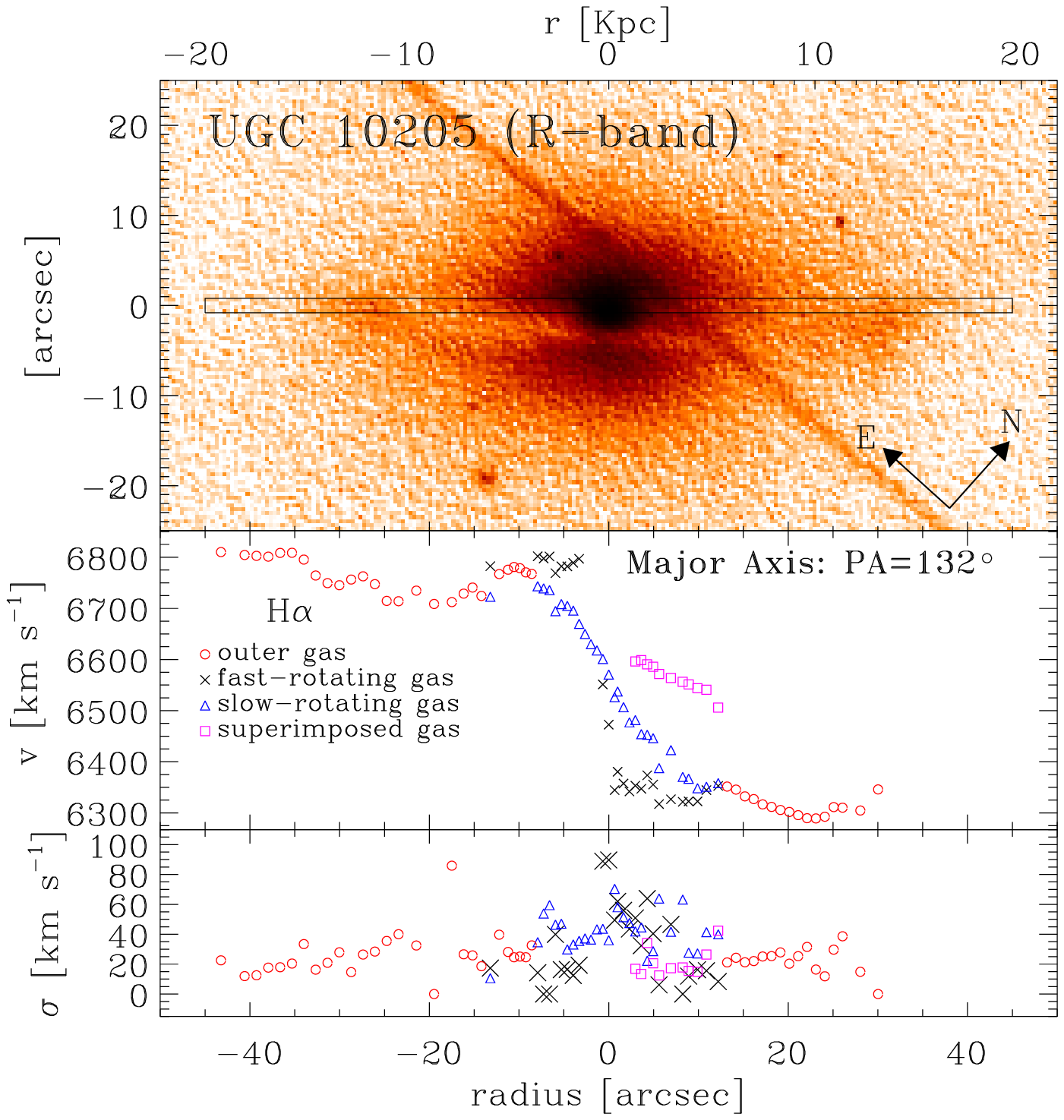}
\includegraphics{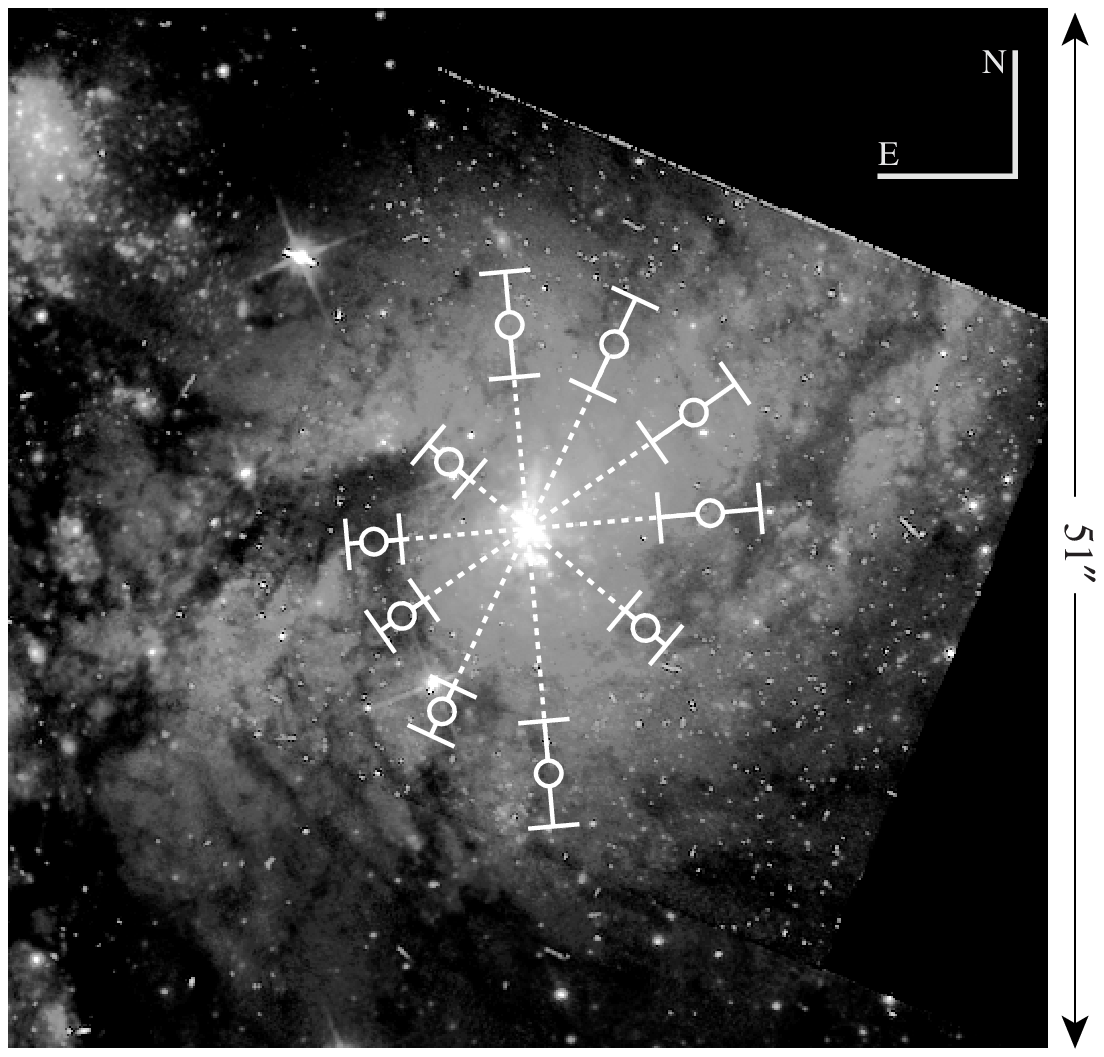}
\vspace*{6.25cm}
\caption[]{\footnotesize
   {\it Left panel}: Above, we can see a {\it R}-band image of the
   edge-on galaxy UGC~10205.  The middle and the bottom windows show
   the velocity curve and the velocity dispersion curve of the
   different gaseous components present in this galaxy.  {\it Right
   panel}: Archival WFPC2 image of NGC~6221 in the F606W filter ({\it
   V}-band).  White lines indicate the five slit position of our
   spectra; location and FWHM of peaks in [NII]/H$\alpha$ for each
   spectrum are noted.}
\label{NGC6221-shocks}
\end{figure}

\section{NGC 4340: Double Bar + Fossil Nuclear Ring}

NGC 4340 is a double-barred SB0 galaxy in the Virgo cluster (Wozniak
et al. 1995).  Using new optical images, we have
found that it also contains a luminous stellar nuclear ring, lying just
outside the inner bar.  The ring is smooth and does not differ in
color from surrounding regions, which suggests it is relatively old
and free of dust and gas: it may be a ``fossil'' remnant of an earlier
gas-rich, star-forming nuclear ring.

The ring is aligned with, but more elliptical than, the outer disk, so
it is intrinsically elliptical.  In deprojection it trails both bars
by 10--$15^{\circ}$ (the two bars are slightly misaligned, which
suggests they are independently rotating).  This is an unusual
orientation, since most hydrodynamic simulations produce leading
(gaseous) nuclear rings within bars.  We use the major-axis velocity
curve of Simien \& Prugniel (1996) and our own
spectrum along the outer-bar major axis to compute approximate
resonance curves.  The resulting $\Omega - \kappa/2$ curve places the
inner bar within the (inner) ILR; it also suggests that the nuclear
ring is at or just inside of the same ILR. (See Erwin et al. 
2000 for details.)

\end{document}